\documentclass[hyper,letterpaper,12pt]{article}
\usepackage{a4wide}
\usepackage{amsmath}
\usepackage[
      colorlinks=true,
      linkcolor=blue,
      urlcolor=blue,
      filecolor=black,
      citecolor=blue,
            ]{hyperref}
\usepackage{color}
\usepackage{graphicx}
\usepackage{amsfonts}
\usepackage{amssymb}
\usepackage{epstopdf}

\def\beq{\begin{equation}}
\def\eeq{\end{equation}}

\begin{document}

\title{\bf \ Holographic Entanglement Entropy in Insulator/Superconductor Transition}

\author{\large
~Rong-Gen Cai\footnote{E-mail: cairg@itp.ac.cn}~,
~~Song He\footnote{E-mail: hesong@itp.ac.cn}~,
~~Li Li\footnote{E-mail: liliphy@itp.ac.cn}~,
~~Yun-Long Zhang\footnote{E-mail: zhangyl@itp.ac.cn}\\
\\
\small State Key Laboratory of Theoretical Physics, Institute of Theoretical Physics, \\
\small Chinese Academy of Sciences, P.O. Box 2735, Beijing 100190,
China
\\}
\date{\small March 29, 2012}
%2011-04-11
\maketitle

\begin{abstract}
\normalsize We investigate the behaviors of entanglement entropy in
the holographical insulator/superconductor phase transition. We
calculate the holographic entanglement entropy for two kinds of
geometry configurations in a completely back-reacted gravitational
background describing the insulator/superconductor phase transition.
The non-monotonic behavior of the entanglement entropy is found in
this system. In the belt geometry case, there exist four phases
characterized by the chemical potential and belt width.
\end{abstract}

\section{ Introduction}

The AdS/CFT
correspondence~\cite{Maldacena:1997re,Gubser:1998bc,Witten:1998qj,Aharony:1999ti}
 provides a powerful theoretical method to study
the strongly coupled systems in various fields of physics. One of
the most widely investigated objects is the holographic
superconductor
(superfluid)~\cite{Gubser:2008px,Hartnoll:2008vx,Hartnoll:2008kx,Gubser:2008zu,Nishioka:2009zj}.
The physical picture is that some gravity background will become
unstable as one tunes some parameter, such as temperature for black
hole and chemical potential for AdS soliton, to develop some kind of
``hair". This condensation of the ``hair" induces the symmetry
breaking, which results in the non-vanishing vacuum expectation
value of the dual operator in the field theory side.

On the other hand, as a measurement of how a given quantum system is
entangled or strongly correlated, the entanglement entropy is also
considered as a useful tool for keeping track of the degrees of
freedom of strongly coupled systems. Dividing a quantum system into
a subsystem $\mathcal{A}$ and its complement, the entanglement
entropy of $\mathcal{A}$ is known as the famous von Neumann entropy
\begin{equation}
S_\mathcal{A}=-Tr_\mathcal{A}(\hat{\rho}_\mathcal{A}\ln\hat{\rho}_\mathcal{A}),
\end{equation}
where $\hat{\rho}_\mathcal{A}$ is the reduced density matrix for
$\mathcal{A}$ by tracing over its complement of the total density
matrix.  However, the calculation of entanglement entropy for a
given system is found to be very difficult except for the case in
$1+1$ dimensions.

Inspired by the Bekenstein-Hawking entropy of black hole and
motivated by the development of AdS/CFT correspondence,
 the authors in Refs.~\cite{Ryu:2006bv,Ryu:2006ef} have presented a proposal to compute the entanglement
 entropy of conformal field theories (CFTs) from the minimal area surface in gravity side. Precisely,
  in an asymptotically anti de-Sitter (AdS) spacetime, consider a slice at constant radial coordinate $r=1/\epsilon$,
  which is dual to a field theory ``living" on this slice with a UV cutoff labeled by the small
   parameter $\epsilon$. the $\epsilon\rightarrow0$ corresponds to the UV limit of the dual field theory,
   which is known as the UV-IR relation~\cite{Susskind:1998dq}. Search for the minimal area
   surface $\gamma_\mathcal{A}$ in the bulk with the same boundary $\partial\mathcal{A}$ of a region $\mathcal{A}$ on
    that slice. The entanglement entropy of $\mathcal{A}$ with its complement is defined as the ``area law":
\begin{equation}\label{law}
S_\mathcal{A}=\frac{\rm Area(\gamma_\mathcal{A})}{4G_N},
\end{equation}
where $G_N$ is the Newton's constant in the bulk. A proof for this
proposal from the basic principle of holographic duality is given in
Ref.~\cite{Fursaev:2006ih} and further properties of the holographic
entanglement entropy are discussed in Ref.~\cite{Hirata:2006jx}.

This proposal provides an elegant and executable way to calculate
entanglement entropy of a strongly coupled system which has a
gravity dual. Since then a lot of works have been carried
out~\cite{Nishioka:2009un,Albash:2011nq,Myers:2012ed,deBoer:2011wk,Hung:2011xb,Nishioka:2006gr,
Klebanov:2007ws,Pakman:2008ui,Ogawa:2011fw} for investigating and
calculating the entanglement entropy in various gravity theories. In
particular, Refs.
\cite{Nishioka:2006gr,Klebanov:2007ws,Pakman:2008ui} presented the
calculations of entanglement entropy for AdS soliton geometry and
Ref. \cite{Ogawa:2011fw} considered the case with higher derivative
corrections. Very recently, the authors of Ref.~\cite{Albash:2012pd}
have studied the entanglement entropy in the holographic
metal/superconductor phase transition. It was found that the
entanglement entropy in superconducting phase is always less than
the one in the metal phase.   The entanglement entropy is
continuous, but does not smoothly cross the second order phase
transition.

 The holographic insulator/superconductor phase
transition was first studied in Ref.~\cite{Nishioka:2009zj}, where
the AdS soliton background \cite{Horowitz:1998ha} was used to mimic
the insulator/superconductor phase transition at zero temperature.
 Motivated by Ref.~\cite{Albash:2012pd} we are going to explore the
behaviors of the entanglement entropy in the
insulator/superconductor phase transition.

The model of the holographic insulator/superconductor phase
transition is constructed in the
 Einstein-Maxwell-scalar theory with a negative cosmological constant in five-dimensional spacetime.
 After completely solving the set of coupled equations of motion, we calculate the entanglement entropy
 for two geometry configurations, one is a half space and the
 other is a belt. In both cases, we find a discontinuity in the slope of the entanglement entropy as a
 function of chemical potential across the phase transition. But, unlike the case of the metal/superconductor phase
 transition~\cite{Albash:2012pd}, after condensate, the entropy first rises as we increase the
  chemical potential $\mu$, arrives at its maximum at a certain chemical potential and then it decreases
  monotonously. For the belt geometry, there exits a so-called ``confinement/deconfinement" transition controlled by
  belt width. Our
results confirm the
  prediction in Ref.~\cite{Klebanov:2007ws} that this phase transition must appear in any
  consistent gravity dual of a confining theory. So there are totally four ``phases" parameterized
  by chemical potential $\mu$ and belt width $\ell$. The complete ``phase diagram" is drawn
  in Figure.\eqref{phasediag}. Here we would like to address that in
  fact there does not exist the parameter ``belt width $\ell$" in the
  dual field theory. When one changes $\ell$, that a new ``phase" occurs
  in the entanglement entropy manifests a length scale of dual
  field theory under consideration. For example, in this paper the
  length scale is nothing, but the correlation length of the
  confining theory.

This paper is organized as follows. In Section
\eqref{sect:background}, we briefly review the holographic
insulor/superconductor phase transition  presented in
Ref.~\cite{Horowitz:2010jq}, and give the complete equations of
motion to be solved. In Section\eqref{sect:conductor}, the fully
  back-reacted system is solved by shooting method. In Section\eqref{sect:entropy} we explore
the behaviors of the entanglement entropy in
insulator/superconductor phase transition. The conclusions and some
discussions are included in Section \eqref{sect:summary}.

%%
%%%%%
%%%section%2%%%%%%%%%%%%%%%%%%%%%%%%
%%%section%2%%%%%%%%%%%%%%%%%%%%%%%
%%%%%%%%%
%_________________________________________________________________%

\section{AdS Soliton Background}
\label{sect:background}

Let us start with the Einstein-Maxwell-complex scalar theory with a
negative cosmological constant in five-dimensional spacetime
\begin{equation}\label{action}
S=\int d^5x\sqrt{-g}\left(\mathcal{R}+\frac{12}{L^2}-\frac{1}{4}F^{\mu\nu}F_{\mu\nu}-|D\psi|^2 -m^2|\psi|^2\right)\;,
\end{equation}
where $L$ is the radius of AdS spacetime, field strength $F=dA$ and
$D_\mu=\nabla_\mu-iqA_\mu$. When the Maxwell field and scalar field
are vanishing, the above action admits an exact solution named as
AdS soliton
\begin{equation}\label{soliton}
ds^2=\frac{L^2dr^2}{r^2f(r)}+r^2(-dt^2+dx^2+dy^2+f(r)d\chi^2),\qquad f(r)=1-\frac{r_0^4}{r^4}\;.
\end{equation}
This asymptotically AdS solution approaches to $R^{1,2}\times S^1$
near the boundary at $r\to \infty$. Moreover, the compactification
$\chi\sim\chi+\pi L/r_0$ needed to avoid a conical singularity at
$r=r_0$ gives a dual picture of a three-dimensional field theory
with a mass gap, which resembles an insulator in the condensed
matter physics~\cite{Nishioka:2009zj}. The geometry in the
$(r,\chi)$ sector just looks like a cigar with its tip at $r=r_0$.
The temperature associated with the soliton background is zero.

In order to mimic the superconductor phase, we need the soliton
solutions with nonvanishing $\psi$. Furthermore, the back reaction
of the matter fields must be taken into account in order to see the
difference of the holographic entanglement entropy. So we choose the
ansatz as follows
\begin{equation}
ds^2 =\frac{dr^2}{r^2B(r)}+r^2\left(-e^{C(r)}dt^2+dx^2+dy^2+e^{A(r)}B(r)d\chi^2\right),\ \
A_t=\phi(r),\ \ \psi=\psi(r).
\end{equation}
Here we have  set $L=1$ without lose of generality. We require that
$B(r)$ vanishes at the tip of the soliton.  And in order to obtain a
smooth geometry at the tip $r_0$, $\chi$ should be made with an
identification
\begin{equation}
\chi\sim\chi+\Gamma,\qquad \Gamma=\frac{4\pi e^{-A(r_0)/2}}{r_0^2B'(r_0)}\;.
\end{equation}
The independent equations of motion under this ansatz are given as
\begin{eqnarray}
&&
\psi''+\left(\frac{5}{r}+\frac{A'}{2}+\frac{B'}{B}+\frac{C'}{2}\right)\psi'+\frac{1}{r^2B}
\left(\frac{e^{-C}q^2\phi^2}{r^2}-m^2\right)\psi=0\;, \\
&&
\phi''+\left(\frac{3}{r}+\frac{A'}{2}+\frac{B'}{B}-\frac{C'}{2}\right)\phi'-\frac{2\psi^2q^2\phi}{r^2B}=0\;,
\\
&&
A'=\frac{2r^2C''+r^2C'^2+4rC'+4r^2\psi'^2-2e^{-C}\phi'^2}{r(6+rC')}\;,
\label{Aeq}
 \\
 &&
C''+\frac{1}{2}C'^2+\left(\frac{5}{r}+\frac{A'}{2}+\frac{B'}{B}\right)C'-\left(\phi'^2+\frac{2q^2
\phi^2\psi^2}{r^2B}\right)\frac{e^{-C}}{r^2} =0,
\\
&&
B'\left(\frac{3}{r}-\frac{C'}{2}\right)+B\left(\psi'^2-\frac{1}{2}
A'C'+\frac{e^{-C}\phi'^2}{2r^2}+\frac{12}{r^2}
\right) \nonumber \\
&&~~~~~~~~~~~~~~~~~~~~~
+\frac{1}{r^2}\left(\frac{e^{-C}q^2\phi^2\psi^2}{r^2}+m^2\psi^2-12\right)
=0\;,
\end{eqnarray}
where a prime denotes  the derivative with respect to $r$. The
matter fields near the boundary $r\rightarrow\infty$ should behave
as
\begin{equation}\label{asypsi}
\psi=\frac{\psi^{(1)}}{r^{\Delta_{-}}}+\frac{\psi^{(2)}}{r^{\Delta_{+}}}+\ldots,
\end{equation}
\begin{equation}\label{asyphi}
\phi=\mu-\frac{\rho}{r^2}+\ldots,
\end{equation}
where $\psi^{(i)}=\langle \hat{O}_{i}\rangle,~i=1,2$, up to a normalization constant, and $\hat{O}_{i}$ are the
corresponding dual operators of $\psi^{(i)}$ in the field theory side. The conformal dimensions of the
operators are $\Delta_{\pm}=2\pm\sqrt{4+m^2}$. $\mu$ and $\rho$ are the corresponding chemical potential and
 charge density in the
dual field theory, respectively.

There are only four independent parameters at the tip, i.e.,
$r_0,\psi(r_0),\phi(r_0),C(r_0)$. However, the above equations of
motion have two useful scaling symmetries~\cite{Horowitz:2010jq}
\begin{equation}\label{scaling1}
r\rightarrow \alpha r,\qquad (\chi,x,y,t)\rightarrow(\chi,x,y,t)/\alpha,\qquad\phi\rightarrow \alpha\phi,
\end{equation}
\begin{equation}\label{scaling2}
C\rightarrow C-2\ln{\beta},\qquad t\rightarrow \beta t,\qquad\phi\rightarrow \phi/\beta.
\end{equation}
These two symmetries allow us to pick any values for the position of
the tip $r_0$ and $C(r_0)$. In our numerical calculations, we will
simply choose $r_0=1,C(r_0)=0$. Furthermore, to recover the pure AdS
boundary, we also need $A(\infty)=0$ and $C(\infty)=0$. In general,
a solution will have a nonvanishing $C(\infty)$. In this case we
will use the scaling symmetry~\eqref{scaling2} to shift $C(\infty)$
to zero.

%__________________________________________________________________

\section{Insulator/Superconductor Phase Transition}
\label{sect:conductor}
From the discussion in Section \ref{sect:background}, for given
$m^2,q,\psi(r_0)$, we can solve the equations of motion for the
 system by choosing $\phi(r_0)$ as a shooting parameter.
After solving the coupled
 equations, we can obtain the condensate $\langle \hat{O}\rangle$ from \eqref{asypsi} and read off the chemical
  potential $\mu$ and charge density $\rho$ through \eqref{asyphi}.

In this paper we only focus on the case $m^2=-\frac{15}{4}$ and $q=2$. It is well known that in
 five-dimensional spacetime, when $-4<m^2<-3$, the scalar field admits two different
 quantizations
  related by a Legendre transform~\cite{Klebanov:1999tb}. $\psi^{(i)}$ can either be identified as a
  source or an expectation value. In this paper we set $\psi^{(1)}=0$ and consider $\psi^{(2)}$ acting as
  the vacuum expectation value of the operator  $\hat{O}_{2}$.

\begin{figure}[h]
\centering
\includegraphics[scale=0.74]{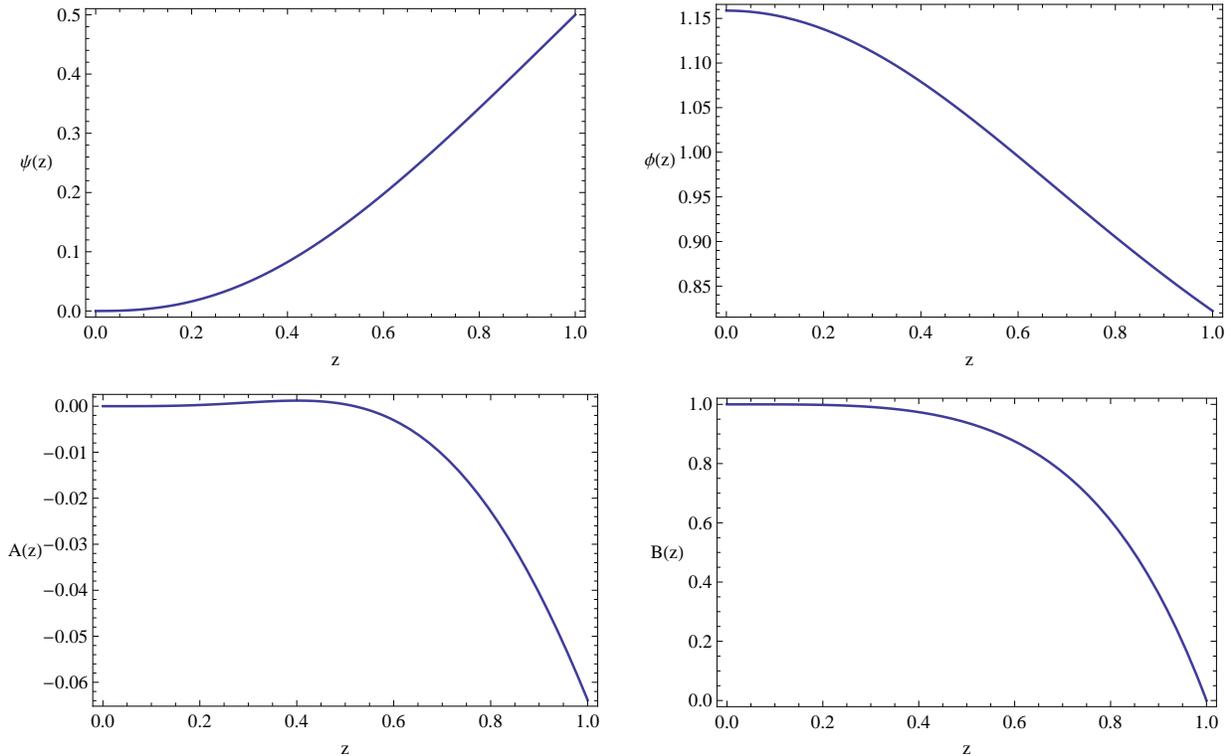}
\caption{\label{function} A typical soliton solution with
nonvanishing scalar hair. Here the value of the scalar at tip is
$\psi_0=0.5$ and the corresponding identification $\Gamma\simeq3.0085$.}
\end{figure}

It is convenient to make a coordinate transformation from
$r-$coordinate to $z-$coordinate by defining $z=1/r$. Therefore, the
infinite boundary is now at $z=0$ and the soliton tip sits at
$z_0=1/r_0=1$. Figure\eqref{function} exhibits a typical solution,
where $A(z)$ and $B(z)$ are two metric functions, and $\psi(z)$ and
$\phi(z)$ are the scalar field and static electric potential,
respectively. Note that the metric functions $A(z)$ and $B(z)$ will
be used in calculating the holographic entanglement entropy.

For different choices of $\psi(r_0)$, the identification length
$\Gamma$ in $\chi$ coordinate will be different. In order to compare
different solutions with the same boundary behavior, we can use the
symmetry \eqref{scaling1} to set the boundary geometry the same.
Taking advantage of the scaling symmetry \eqref{scaling1}, the
relevant quantities scale as follows
\begin{equation}
\Gamma\rightarrow\frac{1}{\alpha}\Gamma,\ \ \mu\rightarrow\alpha\mu,
\ \ \rho\rightarrow\alpha^3\rho,\ \ \langle\hat{O}_2\rangle\rightarrow\alpha^{\frac{5}{2}}\langle\hat{O}_2\rangle.
\end{equation}
The combinations $\mu\Gamma$, $\rho\Gamma^3$ and
$\langle\hat{O}_2\rangle^{\frac{5}{2}}\Gamma$ are all scale
invariant. By using these scale invariant quantities, we can
accomplish our purpose easily. As in our units setup, the
identification length $\Gamma$ in the pure soliton is $\pi$, we
scale all $\Gamma$ for each solution to be $\Gamma=\pi$ from now on.
We should stress here that after the scaling transformation, the tip
$r_0$ will be no longer at $r_0=1$.

\begin{figure}[h]
\centering
\includegraphics[scale=0.92]{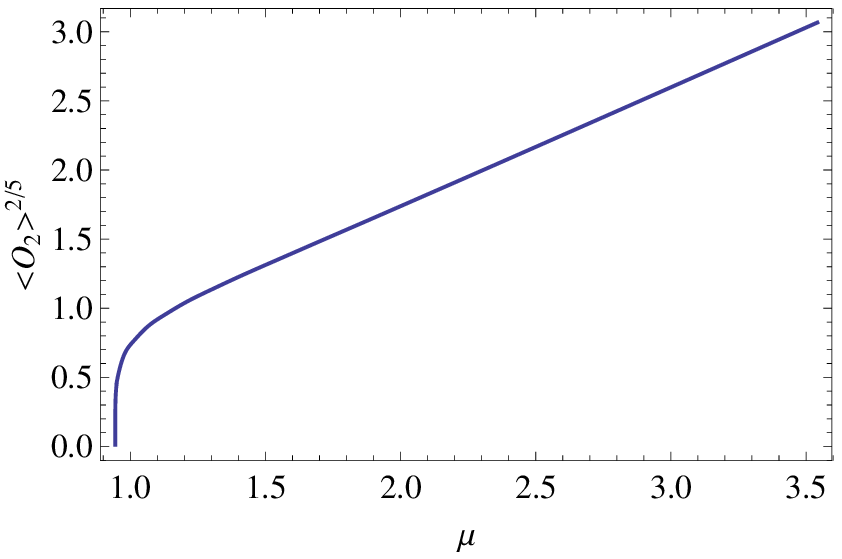}\ \ \ \
\includegraphics[scale=0.88]{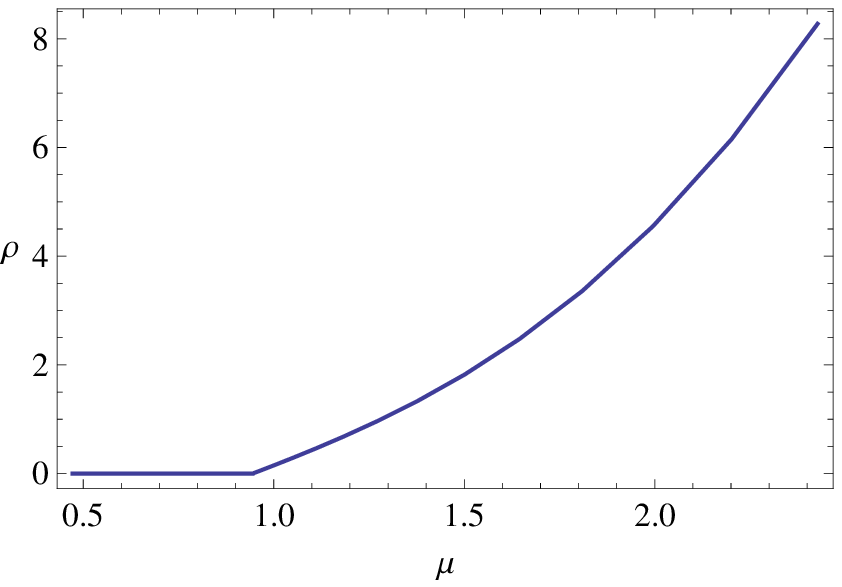} \caption{\label{phase} The condensate of operator
 $\langle\hat{O}_2\rangle$ (left plot) and the charge density $\rho$ (right plot) versus chemical
 potential $\mu$ respectively. Here $m^2=-15/4,\ q=2$ and
  $\Gamma$ is scaled to be $\pi$. The critical chemical potential in this case is $\mu_c\simeq0.9442$.
  It is a typical second order phase transition which can be seen clearly from $\mu\sim\rho$ plot.}
\end{figure}

Figure\eqref{phase} reproduces the results in Ref. \cite{Horowitz:2010jq} by a slightly different manner.
 It shows  the condensate and the charge density as functions of chemical potential.
 It is clear from the figure that as the chemical potential $\mu$ exceeds a critical
  value $\mu_c$ for given mass and charge, the condensation of the operators  emerges.
  This can be identified as a superconductor (superfluid) phase. However, when less than
  $\mu_c$, the scalar field is vanishing and this can be thought of as the insulator phase,
   since this system has a mass gap $\sim 1/\Gamma$ due to the identification in $\chi$ direction.
Note that it is a second order phase transition in our choice of
parameters. In addition,
 let us mention that in the theory  described by the action (\ref{action}),
 except for the AdS soliton solution (insulator) and hairy soliton
 solution (superconducting soliton) mentioned above, there do exist
 another two solutions: the black hole solution without hair (metal
 phase) and black hole solution with hair (another superconducting phase).
 The complete phase diagram has been drawn in
 \cite{Horowitz:2010jq}. The details depends on the coupling constant $q$.
In this paper we just focus on the case with zero temperature
insultor/superconductor phase transition. As shown in
\cite{Horowitz:2010jq}, with our choice of parameters
$m^2=-\frac{15}{4}$ and $q=2$, the phase transition from the
insulator to the superconductor always happens. In particular, for
any values of chemical potential beyond the critical one, the
superconducting phase always exists.

\section{Holographic Entanglement Entropy}
\label{sect:entropy}
%%%%%%%

After finding the gravitational soliton  solution with full back
reaction of matter fields, we are now ready to calculate the
holographic entanglement entropy in this holographic model. Due to
the fact that the choice of the subsystem $\mathcal{A}$ is
arbitrary, we can define infinite entanglement entropies
accordingly. In this paper we first consider a simple case where
$\mathcal{A}$ is chosen to be a half of the total boundary space. We
assume that the subsystem is defined by $x>0$ and extended in the $y$ and
$\chi$ directions, where $-\frac{R}{2}<y<\frac{R}{2}\
(R\rightarrow\infty),\ 0\leq\chi\leq\Gamma$. Then the entanglement
entropy can be deduced from the formula \eqref{law} as
\begin{equation}
S_\mathcal{A}^{half}=\frac{R\Gamma}{4G_N}\int_{r_0}^{\frac{1}{\epsilon}}re^{\frac{A(r)}{2}}dr.
\end{equation}
For the pure AdS soliton solution, $\Gamma=\frac{\pi L}{r_0}$ and we
can read $A(r)$ from metric \eqref{soliton} that $A(r)=0$. So the
entropy can be easily calculated as follows.
\begin{equation}
\label{eq18}
S_\mathcal{A}^{half}=\frac{R\Gamma}{4G_N}\int_{r_0}^{\frac{1}{\epsilon}}rdr=\frac{R\Gamma}{8G_N}r^2\left.
\right |_{r_0}^{1/\epsilon}=\frac{\pi L
R}{8G_Nr_0}(\frac{1}{\epsilon})^2-\frac{\pi
LR}{8G_N}r_0=\frac{R\pi}{8G_N}(\frac{1}{\epsilon^2}-1),
\end{equation}
where $r=\frac{1}{\epsilon}$ is the UV cutoff. Note that in our units setup $L=1,r_0=1,\Gamma=\pi$. The first
 term is divergent ($\epsilon\rightarrow 0$) and represents the ``area law"~\cite{Nishioka:2006gr}. In contrast,
 the second term does not depend on the cutoff and thus is physical important. In fact, this finite term is the
 difference between the entropy in the pure AdS soliton and the one in the pure AdS
 space, because in both cases the divergent part of the entanglement
 entropy is the same. The result in (\ref{eq18}) implies that the entropy in the  AdS soliton is less than the one in
 the pure AdS space.

\begin{figure}[h]
\centering
\includegraphics[scale=0.9]{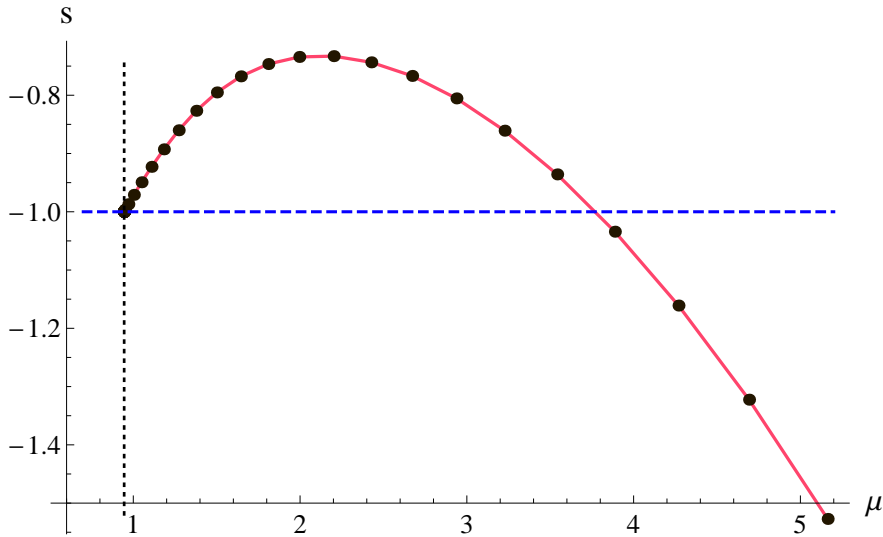}
\caption{\label{halfen} The entanglement entropy as a function of
chemical potential in the half embedding. The solid red curve
denotes the entropy in the superconductor phase, while the dashed
blue line is  for the pure AdS soliton solution. The entropy rises
as the chemical potential $\mu$ is increased  after the phase
transition, arrives at its maximum at a certain $\mu$,  and then
decreases monotonously.}
\end{figure}

 Note that the UV behavior of $S_\mathcal{A}$ will not change after the operator condensation. Thus the divergent
 part of the entropy known as the area law will not change because this part is only sensitive to UV quantities.
 This can be understood from the gravity solution side, the new solution after the operator
 condensation still asymptotically approaches to AdS space near the
 AdS boundary.
  We can write the entanglement entropy in general
in the half embedding case as
\begin{equation}\label{half}
S_\mathcal{A}^{half}=\frac{R\Gamma}{4G_N}\int_{r_0}^{\frac{1}{\epsilon}}re^{\frac{A(r)}{2}}dr=\frac{R\pi}{8G_N}(\frac{1}{\epsilon^2}+s),
\end{equation}
where $s$ has no divergence and $s=-1$ corresponds to the pure AdS
soliton. The value of $s$ as a function of chemical potential $\mu$
in the superconductor phase is presented in Figure~\eqref{halfen}.
One can see from the figure that the entanglement entropy is a
constant in the insulator phase. After condensate, the entropy first
rises and arrives at its maximum as the chemical potential $\mu$
increases, then it decreases monotonously. Thus, from $\mu_c$ to a
certain value of $\mu$, there exits a region where the entropies in
the superconductor phase are larger than the one in the insulator
phase. This behavior of entanglement entropy is quite different from
the one discussed in Ref. \cite{Albash:2012pd}, where the case for
the holographic metal/superconductor phase transition has been studied in
detail. It is shown that the entanglement entropy after condensate
is always lower than the one in the metal phase and decreases
monotonously as  temperature is lowered, and the entropy as a
function of temperate is found to have a discontinuous slop at the
transition temperature $T_c$, which indicates a reduction in the
number of degrees of freedom after condensate. Similarly, we here
also find a discontinuity in the slope of the increasing entropy at
the critical chemical potential $\mu_c$ indicated by the vertical
dotted black line in Figure \eqref{halfen}, which may be considered
as a significant reorganization of the degrees of freedom of the
system. Furthermore, this discontinuity can be regarded as the
signature of the second order phase transition.

We next consider a more nontrivial geometry which is a straight belt
with a finite width $\ell$ along the $x$ direction. The subsystem
$\mathcal{A}$ sites on the slice $r=\frac{1}{\epsilon}$ and expends
in $y$ and $\chi$ directions. The holographic dual surface
$\gamma_\mathcal{A}$ is defined as a three-dimensional surface
\begin{equation}\label{embed}
t=0,\ \ x=x(r),\ \ -\frac{R}{2}<y<\frac{R}{2}\ (R\rightarrow\infty),\ \ 0\leq\chi\leq\Gamma.
\end{equation}
$R$ is the regularized length in $y$ direction. The holographic surface $\gamma_\mathcal{A}$ starts from
$x=\frac{\ell}{2}$ at $r=\frac{1}{\epsilon}$, extends into the bulk until it reaches $r=r_*$,
then returns back to the AdS boundary $r=\frac{1}{\epsilon}$ at $x=-\frac{\ell}{2}$. It is easy to
 obtain the induced metric on $\gamma_\mathcal{A}$ as
\begin{equation}
ds^2 =h_{ij}dx^i dx^j=\left (\frac{1}{r^2B(r)}+r^2\left
(\frac{dx}{dr}\right )^2\right )dr^2+r^2 dy^2+r^2
e^{A(r)}B(r)d\chi^2.
\end{equation}
Using the proposal \eqref{law}, we need to minimize the following
functional
\begin{equation}\label{surface}
S_\mathcal{A}[x]=\frac{R\Gamma}{2G_N}\int_{r_*}^{\frac{1}{\epsilon}}re^{\frac{A(r)}{2}}\sqrt{1+r^4B(r)(dx/dr)^2}dr.
\end{equation}
We can deduce the equation of motion for the minimal surface from \eqref{surface}
\begin{equation}\label{minimal}
\frac{r^5 e^{\frac{A(r)}{2}}B(r)(dx/dr)}{\sqrt{1+r^4B(r)(dx/dr)^2}}=r_s^3 e^{\frac{A(r_s)}{2}}\sqrt{B(r_s)},
\end{equation}
where $r_s$ is a constant.

We are interested in the case that the surface is smooth at $r=r_*$, i.e. $dx/dr|_{r=r_*}\rightarrow\infty$. This condition ensures that $r_s=r_*$. The width $\ell$ of the subsystem $\mathcal{A}$ and $r_*$ are connected by the relation
\begin{equation}\label{width}
\frac{\ell}{2}=\int_{r_*}^{\frac{1}{\epsilon}}\frac{dx}{dr}dr=\int_{r_*}^{\frac{1}{\epsilon}}\frac{1}{r^2\sqrt{B(r)(\frac{r^6B(r) e^{A(r)}}{r_*^6B(r_*)e^{A(r_*)}}-1)}}dr.
\end{equation}
Finally, we obtain the entanglement entropy as
\begin{equation}\label{entropy}
S_\mathcal{A}=\frac{R\Gamma}{2G_N}\int_{r_*}^{\frac{1}{\epsilon}}\frac{r^4\sqrt{B(r)}e^{A(r)}}{\sqrt{r^6B(r) e^{A(r)}-r_*^6 B(r_*) e^{A(r_*)}}}dr=\frac{R\Gamma}{4G_N}(\frac{1}{\epsilon^2}+s),
\end{equation}
where the UV cutoff has been taken into consideration. However,
there is also a disconnected solution describing
 two separated surfaces that are located at $x=\pm\frac{\ell}{2}$, respectively. This configuration is
 just two copies of the half embedding solution that we discussed above. So the entanglement entropy
 of the disconnected configuration is trivially twice of the half embedding case \eqref{half}.
\begin{figure}[h]
\centering
\includegraphics[scale=0.55]{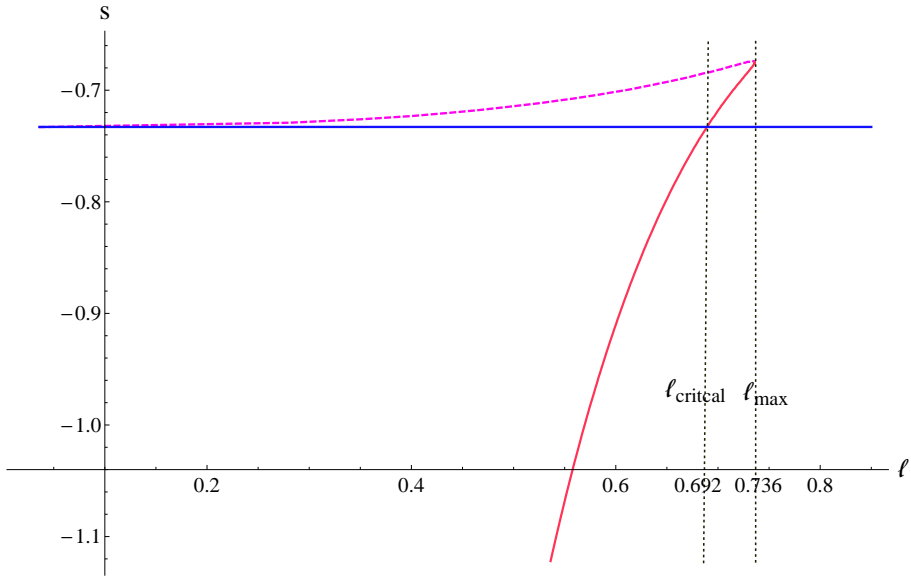}
\caption{\label{Entropy} The entanglement entropy as a function of
strip width $\ell$ for $\mu=2.2016$. The dashed magenta and solid red
curves come from the connected solutions, while the solid blue one
comes from the disconnected one. The lowest curve is physically
favored compared with others because it has minimal entropy.}
\end{figure}

We find the behaviors of the entanglement entropy as a function of
the belt width $\ell$ are quite similar for various chemical
potential $\mu$ or condensation $\langle\hat{O}\rangle$. The result
for $\mu=2.2016$ is shown in Figure~\eqref{Entropy} as a typical
example. When $\ell>\ell_{max}$, there does not exist connected
surface. The physically favored solution is replaced by the trivial
disconnected one in this case. On the other hand, when $\ell
<\ell_{max}$, we have three different branches of entropy for a
given $\ell$ , i.e., the upper connected branch (dashed magenta
curve), the lower connected branch (solid red curve) and the middle
disconnected branch (solid blue line). The physical entropy is
determined by the choice of the lowest one. As we can see from
Figure~\eqref{Entropy}, for small $\ell$, the lower connected branch
is favored. However, there is a critical value $\ell_{critical}$
above which the lower connected branch is not a physical one since
its entropy is larger than the one for the disconnected solution.
Thus, when $\ell$ increases, a phase transition occurs as belt width
$\ell=\ell_{critical}$. This is just the so-called
``confinement/deconfinement" phase transition discussed intensively
in Refs.
\cite{Nishioka:2006gr,Klebanov:2007ws,Pakman:2008ui,Ogawa:2011fw}.
The transition can be understood as the fact that no correlation
could be contributed among a distance larger than $\Gamma$ due to
the mass gap $\sim 1/\Gamma$.
\begin{figure}[h]
\centering
\includegraphics[scale=0.38]{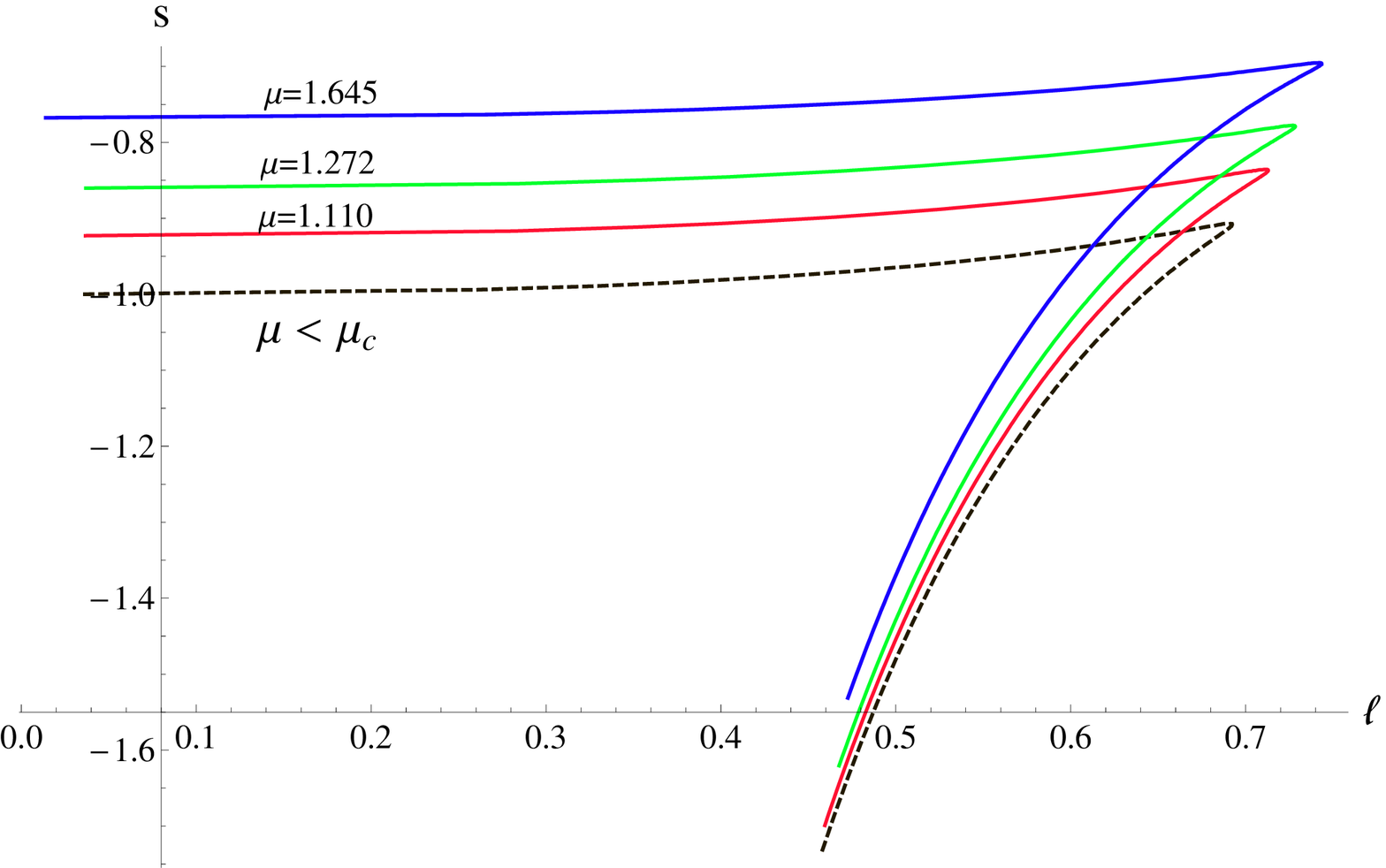}
\includegraphics[scale=0.38]{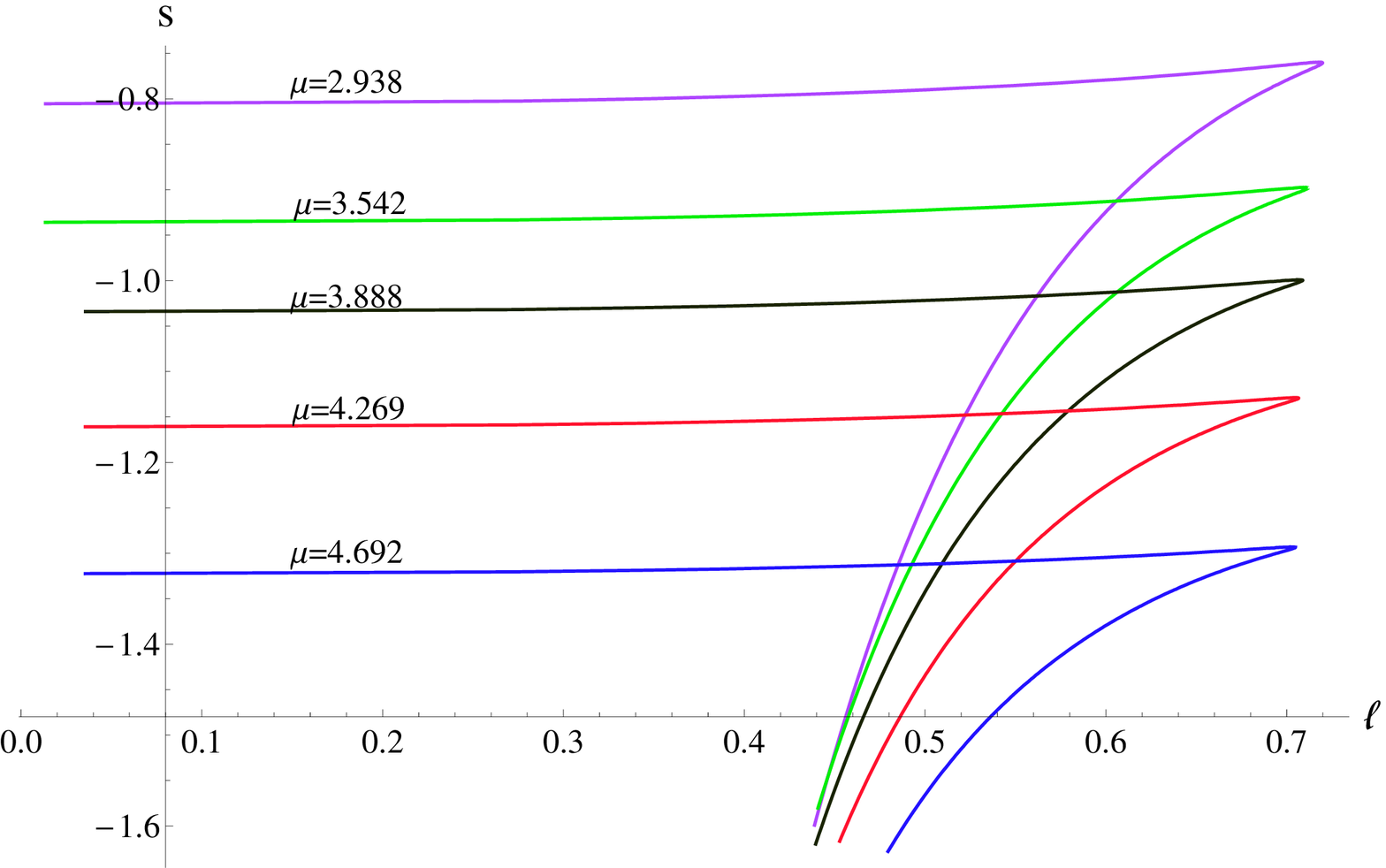}
\caption{\label{little} The entanglement entropy as a function of
belt width $\ell$ for different chemical potential $\mu$. The dotted
black curve is for the pure AdS soliton solution. With a fixed
$\ell$, the entanglement entropy increases (left plot) and then
decreases (right plot) as $\mu$ becomes large.}
\end{figure}

More numerical results for entanglement entropy are exhibited in
Figure~\eqref{little} with different chemical potential. For
$\ell<\ell_{critical}$ at fixed chemical potential, the entropy is
dominated by the connected surface and exhibits a non-trivial
dependence on $\ell$. On the contrary, it is governed by the
disconnected configuration and is independent of $\ell$ for
$\ell>\ell_{critical}$. The fact that the phase transition always
appears at critical length $\ell_{critical}$ is observed for all values of $\mu$ matches  the prediction in
Ref.~\cite{Klebanov:2007ws} that this ``confinement/deconfinemnet" phase transition must be exhibited in any consistent
gravity dual of a confining theory. We find that  $\ell_{max}\simeq(0.220\sim0.237)\Gamma$
and $\ell_{critical}\simeq(0.204\sim0.221)\Gamma$ for various chemical potentials up
to $\mu=5.164$ in our numerical calculation. If the belt width is large
enough ($\ell>0.237\Gamma$), the disconnected branch is always favored for
all values of $\mu$. Thus, the behavior of $s$ as a function of $\mu$ for
fixed $\ell$ is the same as the one shown in Figure~\eqref{halfen}.

\begin{figure}[h]
\centering
\includegraphics[scale=0.92]{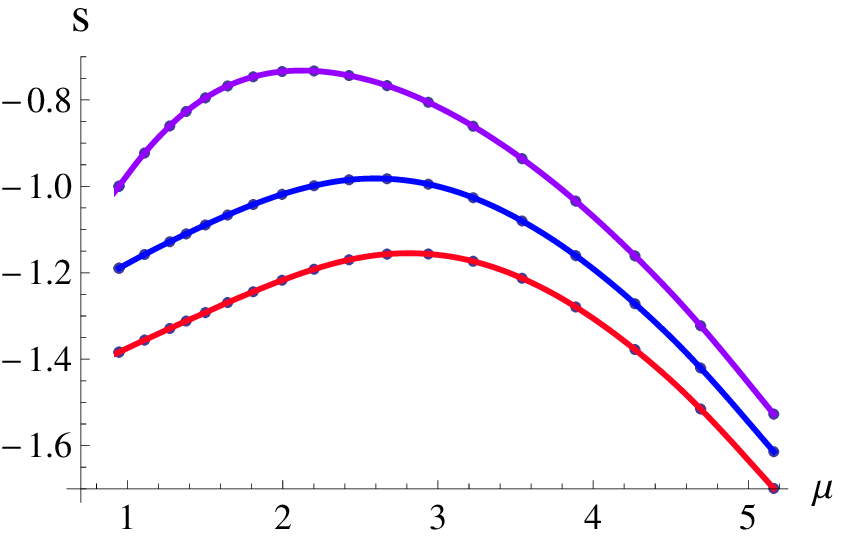}
\includegraphics[scale=0.92]{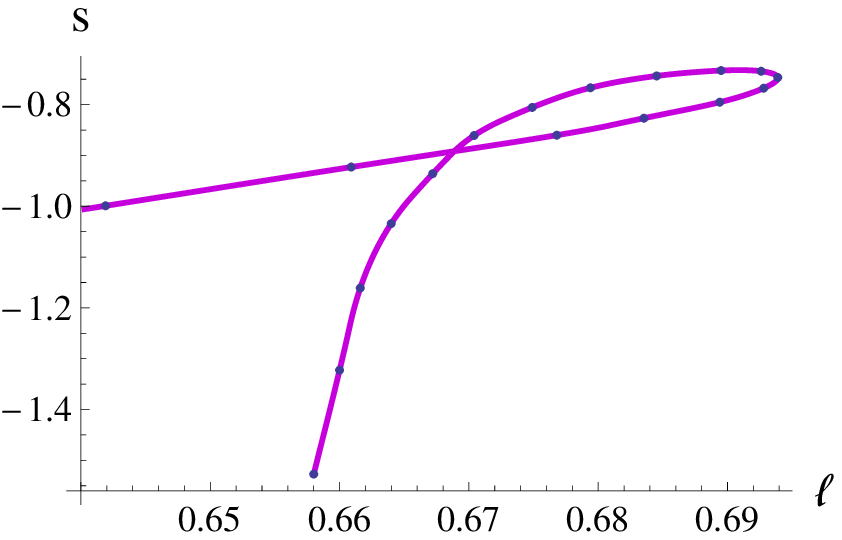}
\caption{\label{critical} {The entanglement entropy of the system
changes with chemical potential $\mu$ for fixing belt width $\ell$
(left plot). The red curve is for $\ell=0.52<\ell_{crtical}$ and the
blue one for $\ell=0.57 <\ell_{critical}$. The purple curve is for
sufficiently large $\ell >\ell_{max}$ in the confining phase. The
 right plot presents the behavior of entropy as a function of belt width at critical point
 of  the ``confinement/deconfinement" transition for different chemical
 potential.}}
\end{figure}

 It is helpful to slice the data in superconducting phase differently. We first study how the entropy of the
 system changes with chemical potential $\mu$ for fixing belt width $\ell$, which is presented in the
 left plot of Figure~\eqref{critical}. It can be seen clearly that the entanglement entropy as a
 function of chemical potential for fixed $\ell$ has similar behavior as in the half embedding case.
 In particular, the purple curve in the confined phase is the same as that in Figure~\eqref{halfen}, in which
 $\ell \to \infty$. As the
 chemical potential $\mu$ increases, the entropy first rises and reaches its maximum at
 a certain value of chemical potential denoted as $\mu_{max}$, then it decreases monotonously. The curve will flatten out as we
  decrease the belt width $\ell$. The curve will finally become a line in the limit $\ell\rightarrow 0$,
   which can be understood from the geometric viewpoint. The minimal surface with small belt length can
   only probe the bulk sufficiently near the boundary $r\rightarrow\infty$ where the metric solution
   is nearly pure AdS. Furthermore, the value of $\mu_{max}$ also decreases as we decrease the belt
    width $\ell$. The values of critical belt width and entropy at the ``confinement/deconfinemnet" transition
     point for different chemical potential are drawn in the right plot of Figure~\eqref{critical}.
      The knot is a consequence of the non-monotonic behavior of the entanglement entropy.

\begin{figure}[h]
\centering
\includegraphics[scale=0.65]{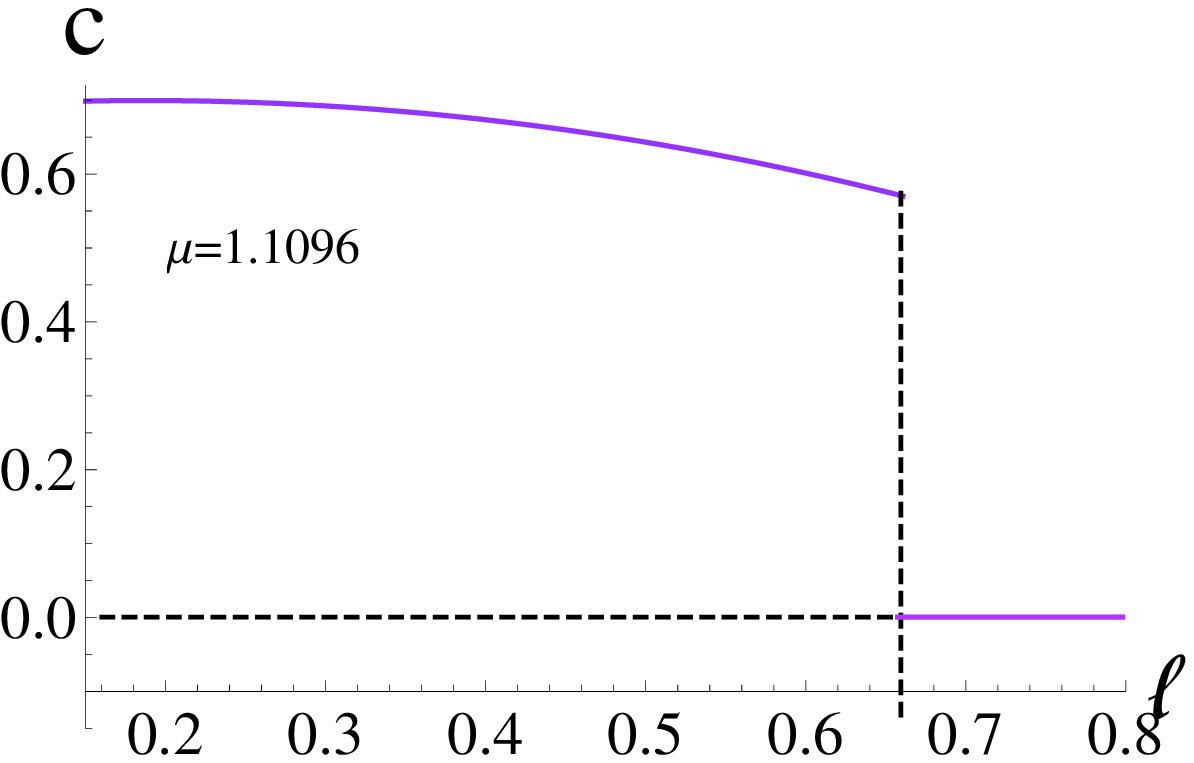}
\includegraphics[scale=0.95]{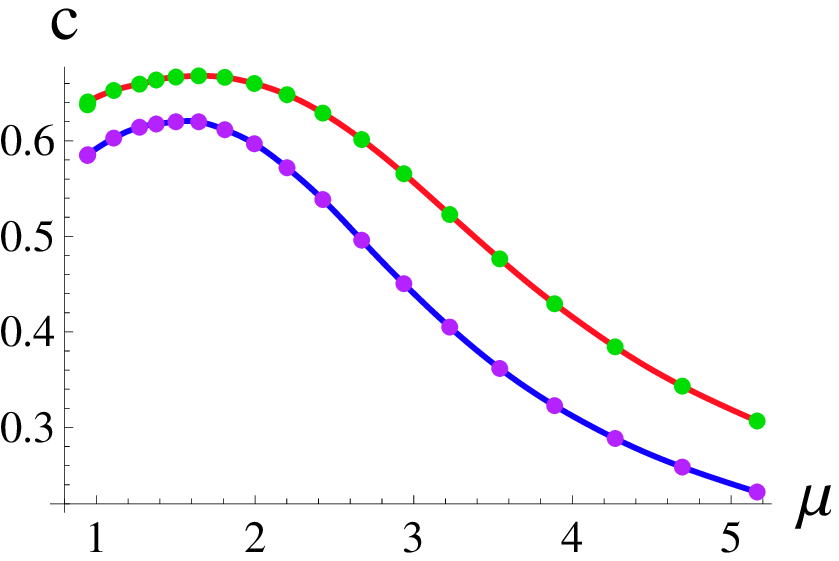}
\caption{\label{cfunc} The entropic c-function as a function of
belt width $\ell$ for $\mu=1.1096$ (left plot) and of chemical
potential $\mu$ at fixed $\ell=0.5$ (right upper curve) and
$\ell=0.6$ (right down curve), respectively.}
\end{figure}

In the deconfinement phase with $\ell<0.221\Gamma$, the connected
branch appears, so the quantitative behavior of $s(\mu,\ell)$
changes. In order to measure the degrees of freedom at energy scale
$\sim1/\ell$ in this situation, we would like to calculate the
so-called entropic c-function first introduced in
Ref.~\cite{Nishioka:2006gr}
\begin{equation}\label{cfun}
c(\ell)=\ell^3\frac{ds(\ell)}{d\ell}.
\end{equation}
The entropic c-function as a function of belt width $\ell$ for a
fixed $\mu$ is plotted in Figure~\eqref{cfunc}. Just as it is
expected, the value of $c$ decreases monotonically as we increase
the belt width $\ell$. Due to the appearance of mass gap
$\sim1/\pi$, the c-function jumps to zero at the critical width
$\ell_{critical}$ ($\ell_{critical}\simeq0.661$ in
Figure~\eqref{cfunc}) and keeps vanishing for larger $\ell$. This is
just the reason that one calls the phase transition mentioned above
the ``confinement/deconfinement" phase
transition~\cite{Nishioka:2006gr,Klebanov:2007ws,Pakman:2008ui,Ogawa:2011fw}.
We also show the change of entropic c-function as we tune the value
of chemical potential for a fixed small $\ell$ in
Figure~\eqref{cfunc}.  We can also see from the figure the
non-monotonic behavior with the increment of chemical potential.
Furthermore, the curve becomes flat as we decrease the belt width
$\ell$. It can be expected that the curve will finally become a line
in the limit $\ell\rightarrow 0$ from the perspective of gravity
side. The behavior of entropic c-function as a function of chemical
potential for a fixed small $\ell$ is quite qualitatively similar to
the entropy in Figure~\eqref{critical}. The key difference here is
that the entropy increases as we increase the belt width, while the
c-function decrease for fixed chemical potential. But an important
feature we observed here is that in both phases, ``confinement and
deconfinement phases", the entanglement entropy is always
non-monotonic in the superconducting phase.

\begin{figure}[h]
\centering
\includegraphics[scale=0.8]{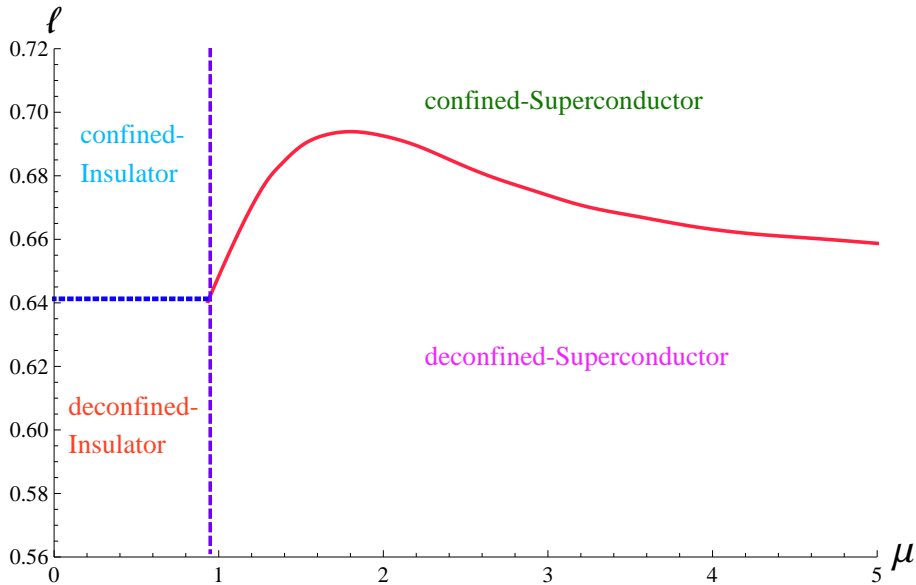}
\caption{\label{phasediag} The ``phase diagram" of entanglement
entropy for a belt geometry in the holographic
insulator/superconductor model for $m^2=-\frac{15}{4}$, $q=2$, and
$\Gamma=\pi$. The phase boundary between the confinement phase and
deconfinement phase is denoted by the dotted blue line and solid red
curve, while the insulator phase and superconductor phase are
separated by the vertical dashed line.}
\end{figure}

To summarize, there do totally exist four phases probed by the
entanglement entropy for a belt geometry for this holographic
system, i.e., the confinement/deconfinement insulator phase, and
confinement/deconfinement superconductor phase.  These phases are
characterized by the chemical potential and belt width. In
particular, the belt width controls the ``confinement/deconfinement"
phase transition. Strictly speaking, the term phase transition here
is inappropriate since the system itself, i.e., the state of the
boundary field theory, does not change at all as one changes the
belt width $\ell$. However, this behavior is reminiscent of many
thermodynamic phase transitions in holographic calculations (see,
for example Refs.~\cite{Chamblin:1999tk,Mateos:2006nu}) and so here
we adopt the terminology ``phase transition" to convey this picture
as in the
literature~\cite{Nishioka:2006gr,Klebanov:2007ws,Pakman:2008ui,Ogawa:2011fw}.
Then one can draw the ``phase diagram" as in
Figure~\eqref{phasediag} for the system. The left part of the
diagram is not new but just the one for the pure AdS soliton case
which has been studied in Ref.~\cite{Nishioka:2006gr}. In the
superconductor phase, we find a non-trivial phase boundary between
the confinement and deconfinement phases. In this sense our work is
a generalization of
Refs.~\cite{Nishioka:2006gr,Klebanov:2007ws,Pakman:2008ui,Ogawa:2011fw}
to the superconducting soliton solution.

The non-monotonic behavior of the entanglement entropy in the
superconductor phase looks strange at first glance. It is quite
different from the case in the metal/superconductor
transition~\cite{Albash:2012pd}, in which the result of the
entanglement entropy decreasing as the temperature is lowered agrees
with the fact that more degrees of freedom will be condensed in
lower temperature.  In our case, one can see from
Figure~\eqref{critical} and Figure~\eqref{cfunc} that the behavior
of both entropy and entropic c-function can be affected by belt
width. However, the belt width $\ell$ does not play the essential
role for the non-monotonic behavior. As we observed above, the
non-monotonic behavior  occurs in both the confinement and
deconfinement superconducting phases.

Due to the lack of the knowledge of the dual field theory in the
``bottom-up" approach, we do not know the details of the microscopic
degrees of freedom and thus it is quite difficult to give a definite
interpretation of the non-monotonic behavior of the entanglement
entropy in the holographic insulator/superconductor phase
transition. In order to understand the numerical result, however, we
here try to give a simple physical picture as follows. Note that the
entanglement entropy as a function of temperature near the critical
temperature $T_c$ shows the same behavior in s-wave and p-wave
holographic metal/superconductor
models~\cite{Albash:2012pd,Cai:2012nm}. The metal phase can be
thought of as the one filled with free charge carriers, such as
electrons. At the critical temperature the condensate turns on and
the free charge carriers are continuously condensed to a certain
microstate as temperature is lowered. As a result, the entanglement
entropy in the superconducting phase is always less than the one in
the metal phase and decreases monotonically as temperature is
lowered. In contrast, the insulator phase can be considered as
having a little degrees of freedom and almost all charge carriers
are bound and can not move freely. This is due to the appearance of
a mass gap $\sim1/\Gamma$ which is the crucial distinction between
the holographic metal/superconductor and insulator/superconductor
transition. At the beginning of the phase transition the condensate
emerges. Thus, a new kind of charge carriers (some quasi-particles
like cooper pairs) that can move unconstrained make contribution to
the degrees of freedom, which leads to the increase of the
entanglement entropy.  When the chemical potential is enough large,
however, all bounded carriers are condensed to form cooper pairs.
The system in this case is full of ``cooper pairs" and thus is an
ordered state seen from momentum space. As a result, the number of
degrees of freedom now are even less than the one in the insulator
phase.  The above argument may account for the non-monotonic
behavior of the entanglement entropy in the insulator/superconductor
phase transition.

To further understand the non-monotonic behavior of the entanglement
entropy in the insulator/superconductor phase transition, it might
be helpful to recall the microscopic model presented in
\cite{Nishioka:2009zj}, where the authors wanted to compare the
holographic insulator/superconductor transition with high-$T_c$
cuprates, in particular with the RVB (resonating valence bond)
scenario of high-$T_c$ electron doped cuprates (e.g.,
Nd$_{2-x}$Ce$_x$CuO$_4$). The phase diagrams for both systems are
quite similar. The high-$T_c$ cuprate at $x = 0$ (in the under-doped
region, here $x$ stands for the electron doping) is the
antiferromagnetic insulator. By doping holes, one frustrates the
antiferromagnetic order and eventually destroys it. Beyond some
critical doping $x_c$, the superconducting ground state emerges. The
physics of the under-doped cuprates is well-described by the t-J
model~\cite{Nishioka:2009zj}. In the slave-boson (SU(2) slave-boson)
approach, the electron operators can be decomposed into bosonic and
fermionc parts, called holons and spinons, respectively.  In the
confined insulator phase, spinons and holons are completely glued
together. At the beginning of phase transition, the holons get
condensed and the degrees of spinons might be released. This might
be the reason that the entanglement entropy increases compared to
the case in the insulator phase. When the electron doping continues
to increase (chemical potential gets large enough in the holographic
insulator/superconudctor phase transition), the spinons will form
cooper pairs and the degrees of freedom associated with spinons get
condensed. In that case, the entanglement entropy of the system is
even less than the case in the insulator phase. This model might
explain the non-monotonic behavior of the entanglement entropy. But
one caveat should be stressed here that in the gravity setup of the
holographic insulator/superconductor model, an extra scalar field
which is dual to the fermion bilinear in the field side does not
included~\cite{Nishioka:2009zj}. But one expects that including the
extra scalar field to the model will not change the background
geometry very much, so that the entanglement entropy keeps the same
behavior. Of course, the above discussions are just some
speculations, further more deep understanding for the non-monotonic
behavior of the entanglement entropy is called for.

\section{Summary and Discussions}
\label{sect:summary}

By using entanglement entropy as a probe, in this paper, we
investigated the holographic insulator/superconductor phase
transition in a five dimensional Einstein-Maxwell-complex scalar
theory with a negative cosmological constant, we calculated the
entanglement entropy of dual field theory with two kinds of geometry
in the AdS boundary: one is a half space, and the other is a belt
geometry with width $\ell$. In the former case, we found that there
does exist a discontinuity of the slop of the entropy with respect
to the chemical potential at the phase transition point. This
discontinuity could be thought of as a significant reorganization of
the degrees of freedom of the system and a signature of the phase
transition. Beyond the critical chemical potential, we found that
the entanglement entropy is not monotonic in the superconductor
phase: at the beginning of the transition, the entropy increases and
arrives at its maximum at a certain chemical potential, and then
decreases monotonically.

In the belt geometry case, the so-called ``confinement/deconfinement
phase transition" happens in both the insulator and superconductor
phases. As $\ell<\ell_{critical}$ the connected surface is favored
and the entropy depends on belt width $\ell$ non-trivially, while as
$\ell>\ell_{critical}$ the disconnected surface is favored and the
entropy is $\ell$ independent. The results enhance the prediction
made in Ref.~\cite{Klebanov:2007ws} that this phase transition must
happen in any consistent gravity dual of a confining theory. Thus,
as a result, there do exist four different ``phases" characterized
by chemical potential $\mu$ and belt width $\ell$, in the
holographic insulator/superconductor phase transition. The complete
``phase diagram" is drawn in Figure~\eqref{phasediag}.

We observed the non-monotonic behavior of the entanglement entropy
with respect to the chemical potential $\mu$ in both the
``confinement/deconfinement superconductor phases. The entanglement
entropy firstly rises and reaches its maximum at a certain chemical
potential and then decreases monotonically (see Figure
(\ref{halfen}) and Figure~\eqref{critical}). Although the
entanglement entropy and entropic c-function depend on the belt
width, it does not play any essential role in the non-monotonic
behavior of the entanglement entropy. The non-monotonic behavior of
the entanglement entropy could be thought of as a key characteristic
of difference between the holographic insulator/superconductor
transition  and the metal/superconductor transition. In the latter
case, the entanglement entropy as a function of temperature near
  the critical transition point always monotonically decreases as
  temperature is lowered~\cite{Albash:2012pd,Cai:2012nm}. Clearly
it would be of some interest to check whether the non-monotonic
behavior of the entanglement entropy is universal in other
holographic insulator/superconductor model, for example, the p-wave
case~\cite{Akhavan:2010bf}.

In addition it is worth stressing here that we did not have a clear
microscopic interpretation for the non-monotonic behavior of the
entanglement entropy, but we did present some arguments which try to
give a physical picture for this behavior. Needless to say, it is
quite necessary to deeply understand this behavior further.

In this paper, we have limited ourselves to the choice of parameters
as $\psi^{(1)}=0$, $m^2=-\frac{15}{4}$, and $q=2$.  Note that the
additional complications
 appear for sufficiently small charge $q$. It is worthy studying the cases with
  other ($m^2,\ q$)'s and the case at finite temperature. It is also interesting to
 study entanglement entropy in other holographic superconductor models, such as the holographic p-wave
 superconductor~\cite{Gubser:2008wv,Akhavan:2010bf,Cai:2010cv,Cai:2010zm}.
  What's more, it is of some interest to see the effect on the entanglement entropy of the
  magnetic field in the holographic superconductors~\cite{Albash:2008eh,Ge:2010aa,Cai:2011tm}.
   We will leave them for future studies.

\section*{Acknowledgements}

RGC thanks T. Takayanagi for helpful correspondences and the authors
thank Zhang-Yu Nie, Hai-Qing Zhang and Shu-Hao Zou for their helpful
discussions and suggestions. This work was supported in part by the
National Natural Science Foundation of China (No.10821504,
No.10975168 and No.11035008), and in part
 by the Ministry of Science and Technology of China under Grant No. 2010CB833004.

%\appendix

\end{document}